# Secret Key Cryptosystem based on Polar Codes over Binary Erasure Channel

R. Hooshmand[1,3], M. Koochak Shooshtari[2], M. R. Aref[3]

[1] Department of Electrical Engineering,
Science and Research Branch, Islamic Azad University, Tehran, Iran
[2] Faculty of Electrical and Computer Engineering
K. N. Toosi University of Technology, Tehran, Iran
[3]Information System and Security Lab (ISSL),
Department of Electrical Engineering, Sharif University of Technology, Tehran, Iran
r.hooshmand@ srbiau.ac.ir, m-koochak@ee.kntu.ac.ir, aref@sharif.edu

*Abstract* — **This manuscript proposes an efficient secret key cryptosystem based on polar codes over Binary Erasure Channel. We introduce a method, for the first time to our knowledge, to hide the generator matrix of the polar codes from an attacker. In fact, our main goal is to achieve secure and reliable communication using finite-length polar codes. The proposed cryptosystem has a significant security advantage against chosen plaintext attacks in comparison with the Rao-Nam cryptosystem. Also, the key length is decreased after applying a new compression algorithm. Moreover, this scheme benefits from high code rate and proper error performance for reliable communication.**

## I. INTRODUCTION

McEliece proposed the first public key cryptosystem based on t-error correcting goppa codes in 1978 [1]. The security of this cryptosystem relies on the fact that the general decoding problem of linear block codes is NP-complete [2]. McEliece cryptosystem has high speed encryption and decryption compared with other public key cryptosystems which are based on number theory such as RSA. However, McEliece cryptosystem has some weaknesses such as large key size and low code rate.

Subsequently, Rao and Nam introduced a secret key cryptosystem based on Hamming codes with high information rate and short code length by keeping the generator matrix secret [3]. The Rao-Nam (RN) cryptosystem is insecure against some known attacks such as chosen plaintext attacks [4]. Many modifications were proposed based to RN secret key cryptosystem by either applying nonlinear codes in the structure of RN cryptosystem or modifying the set of allowed error vectors. However, most of them were shown to be insecure.

In this paper, a new secret key cryptosystem based on polar codes over Binary Erasure Channel with erasure probability $\varepsilon$, BEC($\varepsilon$), is introduced. The proposed cryptosystem is designed to avoid the weaknesses in the RN secret key cryptosystem and is expected to provide more security and efficiency than RN cryptosystem. The rest of the paper is organized as follows. In section 2, we give some basic preliminaries of the polar codes. Section 3 introduces the proposed cryptosystem. Efficiency and Security of the modified scheme are analyzed in sections 4, 5, respectively. Finally, we conclude the paper in section 6.

## II. POLAR CODES

Polar codes, introduced by Arikan, are a class of linear block error correcting codes that can achieve the capacity of symmetric Binary-input Discrete Memoryless Channels (B-DMC) with low encoding/decoding complexity. Consider a B-DMC $W: \mathcal{X} \to \mathcal{Y}$ with binary input alphabet $\mathcal{X} = \{0,1\}$, output alphabet $\mathcal{Y}$ and transition probability function $W(y|x)$ such that $x \in \mathcal{X}$, $y \in \mathcal{Y}$. Let $I(W)$ denote the mutual information with uniform distribution on the inputs. If $W$ is a B-DMC, then $I(W)$ is the symmetric capacity of the channel which measures rate. Let $z(W)$ denote the Bhattacharyya bound of a B-DMC $W$ to measure reliability. These parameters take values in $[0,1]$ and it is expected that $I(W) \approx 1$ iff $z(W) \approx 0$ and $I(W) \approx 0$ iff $z(W) \approx 1$ [5].

In the polar codes, a set of $N = 2^n$ polarized binary-input channels $\{W_N^{(i)}: 1 \leq i \leq N\}$ with channel index "$i$" can be obtained by performing a phenomenon, called channel polarization, on $N$ independent copies of a given B-DMC. This way, symmetric capacity parameters $I(W_N^{(i)})$ of all polarized channels, except for a fraction of them, tend towards 0 or 1 when $N$ is large enough. Channel polarization consists of a channel combining step and a channel splitting step. In the channel combining step, a vector channel $W_N: \mathcal{X}^N \to \mathcal{Y}^N$ is produced in a recursive manner by combining $N$ copies of a given B-DMC $W$. In the channel splitting step, the vector channel $W_N$ splits into $N$ binary input coordinate subchannels $W_N^{(i)}: \mathcal{X} \to \mathcal{Y}^N \times \mathcal{X}^{i-1}, 1 \leq i \leq N$ [5].

Let $A$ be a $K$-element subset of $\{1, 2, \cdots, N\}$ which is called an information set and let $A^c$ be a complementary set of $A$ which is called a frozen set. In polar coding, the information set $A$ is specified such that $I(W_N^{(i)}) \geq I(W_N^{(j)})$ or equivalently $z(W_N^{(i)}) \leq z(W_N^{(j)})$ for all $i \in A$, $j \in A^c$. In other words, it is possible to construct $N$ polarized channels called bit channels such that $N(1 - I(W))$ of them, with channel indices $i \in A^c$, are completely unreliable or noisy and $NI(W)$ of them, with channel indices $i \in A$, are reliable or noiseless [6]. The polarized channels are proper for channel coding by transmitting a sequence of $K$ variable or information bits, $u_A = (u_i, i \in A)$, over the noiseless channels and transmitting a sequence of $N - K$ fixed or frozen bits, $u_{A^c} = (u_i : i \in A^c)$, over the noisy channels [5].

Polar coding is a code construction method that can achieve the capacity of B-DMCs such as the Binary Symmetric Channel (BSC) or BEC. To construct an $(N, K)$ polar code, the $K$ bit-channels with the lowest corresponding Bhattacharyya parameters $\{z(W_N^{(i)}), 1 \leq i \leq K\}$ should be selected. The construction method of the polar codes is efficient over Binary Erasure Channels because the Bhattacharyya parameters can be calculated efficiently with complexity $\mathcal{O}(N)$ for these channels [5]. The generator matrix $G_N(A)$ of an $(N, K)$ polar code over BEC is a $K \times N$ submatrix of $F^{\otimes n} = \begin{bmatrix} 1 & 0 \\ 1 & 1 \end{bmatrix}^{\otimes n}$ for any $N = 2^n, n \geq 1$ and $1 \leq K \leq N$ which is constructed as follows:

1- Compute the vector of Bhattacharyya parameters of $N$ bit channels $Z_N = (z_{N,1}, z_{N,2}, \cdots, z_{N,N})$ through the following recursion for $k = 1, 2, 2^2, \cdots, 2^{n-1}$, starting with erasure probability of BEC, $z_{1,1} = \varepsilon$.

$$z_{2k,j} = \begin{cases} 2z_{k,j} - z_{k,j}^2 & 1 \leq j \leq k \\ z_{k,j-k}^2 & k+1 \leq j \leq 2k \end{cases},$$

2- Construct a permutation $\pi_N = (i_1, \ldots, i_N)$ of the set $\{1, 2, \cdots, N\}$ so that, for any $1 \leq j < k \leq N$, the inequality $z_{N,i_j} \leq z_{N,i_k}$ is true. In fact, $i_1$ and $i_N$ are the indices of bit channels which have minimum and maximum values of Bhattacharyya parameters respectively.

3- The generator matrix $G_N(A)$ is defined as the submatrix of $F^{\otimes n} = \begin{bmatrix} 1 & 0 \\ 1 & 1 \end{bmatrix}^{\otimes n}$ consisting of the rows with indices $i_1, \ldots, i_K$ [7].

In nonsystematic polar coding schemes, an input block $\bar{u} = u_1^N = (u_A, u_{A^c})$ which consists of information bits $u_A$ and frozen bits $u_{A^c}$ is encoded to codeword $\bar{x} = x_1^N = \bar{u}G_N = u_A G_N(A) + u_{A^c} G_N(A^c) = u_A G_N(A) + c$ where $G_N = B_N F^{\otimes n}$ is a generator matrix of size $N$ and $B_N$ is a permutation matrix known as bit reversal. Also, $G_N(A)$ and $G_N(A^c)$ are the submatrices of $G_N$ consisting of the rows with channel indices corresponding to information set $A$ and frozen set $A^c$ respectively. Since $c \triangleq u_{A^c} G_N(A^c)$ is a fixed vector, the encoder which maps $u_A$ to $\bar{x}$ is nonsystematic [9].

The codeword $\bar{x}$ is sent into channels which are obtained by $N$ independent uses of $W$ and the corresponding channel output $\bar{y} = y_1^N$ is received. Polar codes can be decoded by low complexity Successive Cancellation (SC) decoding algorithm. Let $\hat{u}_1^N$ denote an estimate of the input block $u_1^N$. After receiving $\bar{y}$, the bits $\{\hat{u}_i, 1 \leq i \leq N\}$ are estimated successively in the following way:

$$\hat{u}_i = \begin{cases} u_i, & \text{if } i \in A^c \\ h_i(y_1^N, \hat{u}_1^{i-1}) & \text{if } i \in A \end{cases},$$

where decision functions $h_i: \mathcal{Y}^N \times \mathcal{X}^{i-1} \to \mathcal{X}, i \in A$, for all $y_1^N \in \mathcal{Y}^N, \hat{u}_1^{i-1} \in \mathcal{X}^{i-1}$ are as follows,

$$h_i(y_1^N, \hat{u}_1^{i-1}) \triangleq \begin{cases} 0, & \text{if } \frac{w_N^{(i)}(y_1^N, \hat{u}_1^{i-1}|0)}{w_N^{(i)}(y_1^N, \hat{u}_1^{i-1}|1)} \geq 1 \\ 1, & \text{otherwise} \end{cases},$$

In fact, the decoder's goal in the polar codes is to produce an estimate $\hat{u}_1^N$ of $u_1^N$ using the knowledge of information set $A$, frozen vector $u_{A^c}$ and received vector $y_1^N$. We have block errors in the SC decoder if $\hat{u}_1^N \neq u_1^N$ or equivalently if $\hat{u}_A \neq u_A$. The computational complexity of encoding and the complexity of SC decoding are both $\mathcal{O}(N \log N)$. The upper bound on error probability under SC decoding for any B-DMC $W$ and any selection of the parameters $(N, K, A)$ is as follows [5]:

$$P_e(N, K, A) \leq \sum_{i \in A} z(W_N^{(i)}),$$

Also, it is proved that reliable communication for polar codes over BEC with SC decoder is obtained when the following relation is satisfied [10, 11].

$$R < I(W) - N^{-1/\mu}, \tag{1}$$

Where $\mu$ is called scaling exponent and its value for transmission over BEC is $\mu \approx 3.6261$. Indeed, it is a tradeoff between the rate and the block length in the polar codes for a given error probability when we use the SC decoder. In this paper, the largest code rate which satisfies (1) is named by $R_0$. Table 1 shows the error probability variations of the polar codes with $N = 2^{10} = 1024$ over BEC ($\varepsilon = 0.05$) in terms of various code rates. It is clear that error probability decreases significantly for $R < R_0$ under SC decoding.

TABLE I. The upper bound on error probability, $P_{e_1} = \sum_{i=1}^{K} z(W_N^{(i)})$, of the polar codes with $N = 2^{10}$ over BEC($\varepsilon = 0.05$).

| $R$ | $K$ | $P_{e_1}$ |
| --- | --- | --- |
| 0.9 | 922 | $15 \times 10^{-2}$ |
| 0.85 | 870 | $14 \times 10^{-4}$ |
| $R_0 = 0.8$ | 819 | $1.062 \times 10^{-5}$ |
| 0.75 | 768 | $2.892 \times 10^{-8}$ |
| 0.7 | 716 | $2.958 \times 10^{-11}$ |

III. PROPOSED CRYPTOSYSTEM

In the proposed cryptosystem, we consider $(1024, 768)$ polar code with $n = 10$, $R = 0.75$ over BEC with $\varepsilon = 0.05, I(W) = 0.95$ to achieve high security and reliability in the RN secret key cryptosystem. Also, the value of $R_0 = 0.8$ is obtained from inequality (1) as shown in table 1.

## A. Secret Key

The secret key is composed of the set $\{I(\mathcal{S}), \mathcal{IV}_s, \mathcal{S}, \mathcal{P}\}$ which its elements are explained as follows:

1. Let $I(\mathcal{S}) = \{i_{s,1}, \ldots, i_{s,K}\}$ be a set of $K$ secret channel indices which its elements are selected randomly from the $NR_0$ leftmost indices of permutation $\pi_N = (i_1, \ldots, i_N)$. The construction of $\pi_N$ was explained in section 2. Therefore, the secret generator matrix of an $(N, K)$ polar code is defined as the submatrix of $F^{\otimes n}$ consisting of the rows corresponding to the secret indices $i_{s,1}, \ldots, i_{s,K}$. It is possible that such selection is not the best to achieve channel capacity but in this way, the generator matrix of the polar code is obscured properly from an attacker, as we will see in section 5. In fact, it is a tradeoff between security and efficiency that is almost inevitable in designing of the code based cryptosystems.

2. Let $\mathcal{IV}_s$ be a secret $(N-K)$-bit initial value (seed) of a Linear Feedback Shift Register, LFSR, to generate a sequence of $2^{N-K}$ pseudorandom syndromes synchronously [12]. In this scheme, an $(N-K)$-bit pseudorandom syndrome $s \in F_2^{N-K}$ is considered as the frozen vector $u_{A^c}$. In encryption process of each plaintext block, the pseudorandom syndrome $s = u_{A^c}$ which is employed by the sender must be known to the receiver synchronously.

3. Let $\mathcal{S}_{K \times K}$ be a regular sparse nonsingular scrambling matrix formed by $k_0^2$ binary circulant $l \times l$ submatrices $S_{j,k}, j = 1, \cdots, k_0, k = 1, \cdots, k_0$ over $GF(2)$ with row/column Hamming weight $\mu_S = 2$ such that $K = k_0 l$ [14],

$$\mathcal{S} = \begin{bmatrix} S_{1,1} & S_{1,2} & \cdots & S_{1,k_0} \\ S_{2,1} & S_{2,2} & \cdots & S_{2,k_0} \\ \vdots & \vdots & \ddots & \vdots \\ S_{k_0,1} & S_{k_0,2} & \cdots & S_{k_0,k_0} \end{bmatrix}.$$

4. Let $\mathcal{P}_{N \times N}$ be an $N \times N$ block diagonal permutation matrix formed by $n_0 \times n_0$ submatrices $P_{l \times l}$ over $GF(2)$ such that $N = n_0 l$,

$$\mathcal{P} = \begin{bmatrix} P_{1,1} & 0 & \cdots & 0 \\ 0 & P_{2,2} & \cdots & 0 \\ \vdots & \vdots & \ddots & \vdots \\ 0 & 0 & \cdots & P_{n_0,n_0} \end{bmatrix}.$$

The diagonal elements are circulant permutation submatrices and the other elements are zero submatrices in a way that the Hamming weight of each row or column is one [14]. In the proposed cryptosystem, we have $n_0 = 8$, $k_0 = 6$, $l = 128$ for the (1024,768) polar code.

## B. Encryption

1. The sender first randomly chooses a code in a family of equivalent $(N, K)$ polar codes over BEC($\varepsilon$) by selecting $K$ channel indices randomly from the $NR_0$ leftmost indices of permutation $\pi_N$. Then, the generator matrix $G_N(A)$ is constructed in a similar way discussed in section 2. The generator matrix is defined as the submatrix of $F^{\otimes n}$ consisting of the rows corresponding to the selected secret channel indices. Also, the authorized transmitter and receiver consider the set of selected channel indices, $I(\mathcal{S})$, as an element of secret key set. Then, the sender generates the pseudorandom syndrome $s = u_{A^c}$ for each plaintext block using the LFSR and the secret initial value (seed) $\mathcal{IV}_s$

2. Finally, the plaintext is divided into $K$-bit blocks $M = (m_1, m_2, \cdots, m_K)$ and is encrypted as follows.

$$C = (M\mathcal{S} G_N(A) + u_{A^c} G_N(A^c))\mathcal{P} = MG' + e\mathcal{P},$$

where $C = (c_1, c_2, \cdots, c_N)$ is an $N$-bit ciphertext block, $e = u_{A^c} G_N(A^c) = sG_N(A^c)$ is considered as an $N$-bit perturbation error vector and $G' = \mathcal{S} G_N(A)\mathcal{P} = [g'_{ij}], i = 1, 2, \cdots K, j = 1, 2, \cdots, N$ is an encryption matrix.

## C. Decryption

The authorized receiver decrypts the received vector $r = C + e_{ch} = MG' + e\mathcal{P} + e_{ch}$ which is influenced by channel error $e_{ch}$. The decryption process is performed as below.

1. The receiver applies $\mathcal{P}^T$ to vector $r$ and computes $r' = r\mathcal{P}^T = M\mathcal{S} G_N(A) + u_{A^c} G_N(A^c) + e_{ch}\mathcal{P}^T$. In this case, $e_{ch}\mathcal{P}^T$ is a vector having the same Hamming weight as $e_{ch}$.

2. By using the secret frozen vector $u_{A^c}$ and the set of $K$ secret channel indices $I(\mathcal{S})$, the receiver eliminates $e_{ch}\mathcal{P}^T$ and estimates $K$-bit vector $M' = M\mathcal{S}$ under SC decoding algorithm of the polar codes. At last, $M = M'\mathcal{S}^{-1}$ is obtained.

## IV. EFFICIENCY

To measure the efficiency of the proposed cryptosystem, we consider three factors: Error performance, key length and computational complexity.

### A. Error Performance

In the proposed cryptosystem, we have reliable communication because; the secret channel indices $i_{s,1}, \ldots, i_{s,K}$ are chosen randomly from the best $NR_0$ channel indices corresponding to the noiseless channels. Furthermore, the dimension of the used polar code and the parameters of the BEC are chosen in a way which satisfies (1). In this case, the upper bounds on error probability can be varied from $P_e \leq P_{e_1} = \sum_{i=1}^{K} z(W_N^{(i)})$ to $P_e \leq P_{e_2} = \sum_{i=NR_0-K+1}^{NR_0} z(W_N^{(i)})$ depending on the random selection of the secret channel indices. In the proposed scheme, the upper bounds on error probability can be varied from $P_e \leq P_{e_1} = \sum_{i=1}^{768} z(W_{1024}^{(i)}) \approx 2.892 \times 10^{-8}$ to $P_e \leq P_{e_2} = \sum_{i=52}^{819} z(W_{1024}^{(i)}) \approx 1.062 \times 10^{-5}$.

The variations of the upper bounds on error probability for polar codes with $N = 2^{10}$ over BEC(0.05) in terms of random selection of the set $I(\mathcal{S})$ and code rates $R < R_0$ are shown in table 2. It is clear that unlike the upper bound $P_{e_1}$, the value of $P_{e_2}$ is approximately invariable and is equal to $\sum_{i=1}^{NR_0} z(W_N^{(i)})$.

TABLE II. The variations of the upper bounds $P_{e_1}$ and $P_{e_2}$ for the polar codes with $N = 2^{10}$, $R < R_0$ over BEC(0.05).

| R | K | $P_{e_1}$ | $P_{e_2}$ |
|---|---|---|---|
| 0.75 | 768 | $2.892 \times 10^{-8}$ | $1.062 \times 10^{-5}$ |
| 0.7 | 717 | $2.958 \times 10^{-11}$ | $1.062 \times 10^{-5}$ |
| 0.65 | 665 | $1.2 \times 10^{-13}$ | $1.062 \times 10^{-5}$ |
| 0.6 | 615 | $7.881 \times 10^{-17}$ | $1.062 \times 10^{-5}$ |

*B. Key Length*

The key size of the proposed cryptosystem is computed before and after executing key compression/decompression algorithms which are introduced in [14]. These algorithms are based on circulant block of submatrices in the structure of $S$ and $\mathcal{P}$ matrices. Without executing compression algorithms, we require to store $K$ secret indices $I(s) = (i_{s,1}, \dots, i_{s,K})$ instead of saving the generator matrix $G_N(A)$. Therefore, the upper bound of the required memory for storing the set $I(s)$ is $\mathcal{M}_{I(s)} \leq 11K$ bits. The required memory for storing the initial value of an $(N - K)$-bit LFSR is $\mathcal{M}_{\mathcal{IV}_s} = N - K = 2l$ bits. Furthermore, we require only the first rows of $l \times l$ submatrices $S_{j,k}$, $j = 1, \dots, k_0, k = 1, \dots, k_0$ to store the sparse nonsingular scrambling matrix $S_{K \times K}$. So, the required memory for saving matrix $S$ is $\mathcal{M}_S = k_0^2 l$ bits. Also, the required memory for storing permutation matrices $\mathcal{P}_{N \times N}$ which consists of $n_0$ circulant permutation submatrices $P_{l \times l}$ on its diagonal is $\mathcal{M}_\mathcal{P} = n_0 l$ bits. Therefore; the actual key length of the proposed cryptosystem is computed as follows:

$$\mathcal{M}_{\mathcal{K}_{actual}} = \mathcal{M}_{I(s)} + \mathcal{M}_{\mathcal{IV}_s} + \mathcal{M}_S + \mathcal{M}_\mathcal{P}$$
$$\leq (k_0^2 + 11k_0 + n_0 + 2)l \approx 14.33 kbit.$$

Here, we use compression algorithm $\mathcal{A}$ and decompression algorithm $\mathcal{B}$ for the nonsingular matrix $S$ respectively to reduce the key length [14]. Similar algorithms can be applied to the permutation matrix $\mathcal{P}$.

Algorithm $\mathcal{A}$

Input:
- Sparse scrambling nonsingular matrix $S$.

Output:
- Compressed vector $S_c$.

Algorithm:
1. Consider a full zero vector $S_c$ consisting of $\mu_S k_0^2$ coordinates.
2. for $j = 1$ to $k_0$ do
3.    for $k = 1$ to $k_0$ do
4.       Select nonzero positions in the first row of submatrix $S_{j,k}$ from matrix $S$.
5.       Insert the selected positions from left to right in vector $S_c$.
6.    end for
7. end for
8. return $S_c$.

In the proposed cryptosystem, the Hamming weight of each row/column of $S_{j,k}$ submatrices is $\mu_S$. So, the compressed vector $S_c$ consists of $\mu_S k_0^2$ nonzero positions which involves at most $\mathcal{M}_{S_c} = 8\mu_S k_0^2$ bits of memory. For compressing the matrices $S$ and $\mathcal{P}$, the sender should send a new characteristic vector $CHR = (l, n_0, k_0, \mu_S)$ to the authorized receiver as an element of the secret key. So, by using the compression algorithms, the maximum required memory for saving the compressed secret key is computed as follows:

$$\mathcal{M}_{\mathcal{K}_{comp.}} = \mathcal{M}_{I(s)} + \mathcal{M}_{\mathcal{IV}_s} + \mathcal{M}_{S_c} + \mathcal{M}_{\mathcal{P}_c} + \mathcal{M}_{CHR}$$
$$\leq k_0(11l + 16k_0) + 8n_0 + 2l \approx 9.34 kbit.$$

It is clear that the key length of the proposed cryptosystem is decreased by 35 percent after applying the corresponding compression algorithms to $S$ and $\mathcal{P}$. The intended receiver can decompress vectors $\mathcal{P}_c$ and $S_c$ to obtain matrices $\mathcal{P}$ and $S$, using the secret characteristic vector $CHR$ and the proposed decompressing algorithms. Here, we present decompressing algorithm for $S_c$, similar algorithm can be applied to $\mathcal{P}_c$.

Algorithm $\mathcal{B}$

Input:
- $S_c$, $CHR$.

Output:
- $S$.

Algorithm:
1. Construct a full zero $S_{K \times K}$ matrix consisting of $k_0 \times k_0$ submatrices $S_{l \times l}$;
2. Let $g \leftarrow 1$;
3. for $j = 1$ to $k_0$ do
4.    for $k = 1$ to $k_0$ do
5.       Select, the $g^{th}$ $\mu_S$ coordinates of $S_c$, from left to right.
6.       Insert '1's in the $\mu_S$ positions of the first row of the $S_{j,k}$ (the $(j,k)^{th}$ submatrix of $S$) corresponding to the values of the selected $g^{th}$ $\mu_S$ coordinates.
7.       for $f = 1$ to $l - 1$ do
8.          Shift the first row of $S_{j,k}$, $f$ positions to the right.
9.          Insert $f^{th}$ shift of the first row in $(f + 1)^{th}$ row of $S_{j,k}$.
10.      end for
11.   end for
12. Let $g \leftarrow g + 1$.
13. end for
14. return $S$.

The key length of the proposed cryptosystem is compared with the key length of the Rao-Nam cryptosystem in table 3.

TABLE III. Comparing the Key length of the Rao-Nam and the Proposed cryptosystems.

| Cryptosystem | Rao-Nam | Proposed |
|---|---|---|
| code | Hamming | polar |
| $(N, K)$ | (72,64) | (1024, 768) |
| Rate | $\approx 0.89$ | 0.75 |
| Key length | $\approx 18$ kbits [15] | Before Comp. $\approx 14.33$ kbits<br>After Comp. $\approx 9.34$ kbits |

Although the dimension of used polar code is much larger than the code length of the RN cryptosystem, the compressed key length of the proposed cryptosystem is about 48 percent shorter than the key length of the RN cryptosystem.

## C. Computational Complexity

The Encryption/Encoding complexity can be expressed as follows [14],

$$C_{Enc} = C_{mul}(MG') + C_{mul}(e\mathcal{P}),$$

where $C_{mul}(MG') \leq NK$ is the number of binary operations for multiplying $K$-bit vector $M$ to encryption matrix $G'$. Also, $C_{mul}(e\mathcal{P}) = N$ is the number of required binary operations for multiplying $N$-bit perturbation error vector $e$ to the permutation matrix $\mathcal{P}$. The Decoding/Decryption complexity can be calculated as follows,

$$C_{Dec} = C_{mul}(r\mathcal{P}^T) + C_{SC} + C_{mul}(M'S^{-1}),$$

where the number of binary operations for multiplying $N$-bit received vector to the transposed permutation matrix $\mathcal{P}$ is computed as $C_{mul}(r\mathcal{P}^T) = N$. The complexity of successive cancelation decoding of the polar code is $C_{SC} = \mathcal{O}(NlogN)$. Furthermore, the number of required binary operations for multiplying the k-bit vector $M'$ to the inverse matrix $S^{-1}$ is obtained as $C_{mul}(M'S^{-1}) \leq K^2$.

## V. SECURITY

In this section, we consider some attacks such as the Brute Force, the Rao-Nam, the Struik-Tilburg attacks which have been proposed against the RN cryptosystem.

### A. The Brute Force Attack

In the Brute Force attack, all possible keys are checked systematically until the correct key is found. In fact, this attack is impossible, only if the size of the key space is large enough. In the proposed cryptosystem, the parameters of the key set $\{I(\mathcal{S}), \mathcal{IV}_s, \mathcal{S}, \mathcal{P}\}$ are computed as follows:

i. Since the transmitter randomly selects $K$ indices from the $NR_0$ leftmost indices of permutation $\pi_N$, the number of equivalent polar codes is computed as follows:

$$\mathcal{N}_C(N,K) = \binom{NR_0}{K} = (NR_0)!/(K!(NR_0 - K)!),$$

So the involved code parameters produce a family of $(1024, 768)$ equivalent polar codes over BEC(0.05), which is $\mathcal{N}_C(1024, 768) \approx 2^{271}$. Therefore, there is large enough equivalent polar codes to resist against the Brute Force attack.

ii. The number of perturbation vectors $e = sG_N(A^c) = u_{A^c}G_N(A^c)$ is equal to the number of $(N-K)$-bit pseudorandom syndromes $\mathcal{N}_e = \mathcal{N}_s = 2^{N-K} = 2^{256}$. Hence, finding the perturbation error vectors is infeasible.

iii. The number of nonsingular scrambling matrices $\mathcal{S}_{K \times K}$ over $GF(2)$ is as follows [3],

$$\mathcal{N}_\mathcal{S} = \prod_{i=0}^{K-1}(2^K - 2^i) > 2^{K^2 - K}.$$

In our scheme, for $K = 768$ we have $\mathcal{N}_\mathcal{S} \gg 2^{80}$ which indicates an impractical preliminary work for an attacker.

iv. The number of block diagonal permutation matrices satisfies $\mathcal{N}_\mathcal{P} = (l!)^{n_0}$. In the proposed cryptosystem, the number of these matrices for $l = 128$, $n_0 = 8$, is $\mathcal{N}_\mathcal{P} = (128!)^8 \gg 2^{80}$. So, finding the permutation matrix is infeasible in polynomial time.

### B. The Rao-Nam Attack

The Rao-Nam (RN) attack is a chosen plaintext attack which takes place in two steps as follows [3]:
- The encryption matrix $G'$ is solved from a large set of plaintext-ciphertext $(M, C)$ pairs.
- The plaintext $M$ is obtained from $C$ using $G'$ obtained in the previous step.

Let $M_1$ and $M_2$ be two plaintext vectors which differ only in the $i^{th}$, $i = 1, 2, \cdots, K$ position. Let $C_1 = M_1G' + e_1\mathcal{P}$ and $C_2 = M_2G' + e_2\mathcal{P}$ be the corresponding ciphertext vectors whose difference is computed as follows:

$$C_1 - C_2 = (M_1 - M_2)G' + (e_1 - e_2)\mathcal{P} = g'_i + (e_1 - e_2)\mathcal{P}, \quad (2)$$

where $g'_i$ is the $i^{th}$ row vector of the encryption matrix $G'$. Let $C_j = MG' + e_j\mathcal{P}$ and $C_k = MG' + e_k\mathcal{P}$ be two distinct ciphertexts which are obtained from the same plaintext $M$ whose difference is $C_j - C_k = (e_j - e_k)\mathcal{P}$. The $i^{th}$ row vector of the encryption matrix $G'$ is given by $g'_i = C_1 - C_2 - (e_1 - e_2)\mathcal{P}$. Therefore, every value of $(e_j - e_k)\mathcal{P}$ should be tested for $(e_1 - e_2)\mathcal{P}$ of (2) to obtain $g'_i$. This step must be repeated until all possible pairs of error vectors are tested. The number of distinct error vectors is $\mathcal{N}_e = 2^{N-K}$ and the number of possible values of $(e_j - e_k)\mathcal{P}$ is $\frac{(\mathcal{N}_e^2 - \mathcal{N}_e)}{2}$. In this way, the complete solution of encryption matrix $G'$ must be obtained and verified, because the correctness of each $g'_i$ cannot be verified independently. The work factor of determining the encryption matrix $G'$ from RN attack is $WF = \Omega(2^{(N-K)K})$ for $\mathcal{N}_e = 2^{N-K}$ [3]. Obviously, this attack is infeasible for the proposed cryptosystem with $\mathcal{N}_e = 2^{256}$.

Also, according to [3], the Hamming weight of error vectors $e$ should be approximately $N/2$ to resist against Majority Voting (MV) attack. In [16], Meijer and Tilburg introduced Extended Majority Voting (EMV) attack which can be considered as the generalized the MV attack. It was shown that the RN cryptosystem is vulnerable to the EMV attack because of imposing Hamming weight constraint. The EMV attack is essentially optimal if the Hamming weight of error vectors is equal to $N/2$. Therefore, in order to prevent the EMV attack, the error vectors have to be chosen randomly without any Hamming weight constraints [16]. In the proposed cryptosystem, there are no constraints on the Hamming weight of the perturbation error vectors to avert the EMV attack.

## C. The Struik-Tilburg Attack

The Struik-Tilburg (ST) attack is also a chosen plaintext attack that is described as follows [4]. First, an attacker needs to encipher arbitrary message $M$ until all $\mathcal{N}_e$ distinct ciphertexts $\mathsf{C} = \{C_j = MG' + e_j\mathcal{P}\}_{j=1}^{\mathcal{N}_e}$ are obtained. Then, a directed labeled graph $\Gamma = (\mathsf{C}, E_\Delta^\mathcal{P})$ is constructed by cryptanalyst where each vertex of $\Gamma$ is named by the corresponding ciphertext in $\mathsf{C}$ and each edge from vertex $C_i$ to vertex $C_j$ is labeled as follows.

$$C_i - C_j = (MG' + e_i\mathcal{P}) - (MG' + e_j\mathcal{P}) = e_{i,j}\mathcal{P},$$

A set of permuted error vector differences is denoted by $E_\Delta^\mathcal{P} = \{e_{i,j}\mathcal{P}\}_{i,j=1}^N$. Subsequently, the attacker constructs an automorphism group $Aut(\Gamma)$ so that the edges are invariant through all permutations on $\mathsf{C}$. Then, message $M_i = M + u_i, 1 \leq i \leq K$, is chosen by the attacker where $u_i$ is the unit vector with one '1' in the $i^{\text{th}}$ position. Graph $\Gamma_i = (\mathsf{C}_i, E_\Delta^\mathcal{P})$ is constructed for the message $M_i$ by applying same procedure. Also, the automorphism $\Phi$ is selected randomly from $Aut(\Gamma)$ and $\mathcal{N}_e$ ciphertexts $C_{i,1}, C_{i,2}, \cdots, C_{i,\mathcal{N}_e}$ are obtained by map $\Phi: \Gamma_i \to \Gamma$ in a similar way with $C_1, C_2, \cdots, C_{\mathcal{N}_e}$. The difference $C_{i,1} - C_1$ is calculated as follows,

$$C_{i,1} - C_1 = (M_i G' + e_{i,1}\mathcal{P}) - (MG' + e_1\mathcal{P}) = g_i' + \tilde{e}_{i,1}$$

By this method, the $g_i'$ can be estimated with probability $|Aut(\Gamma)|^{-1}$ because, there exists an automorphism $\Phi$ for which $\tilde{e}_{i,1} = 0$. On the average, the attacker should construct $|Aut(\Gamma)|^K$ encryption matrices $G'$ before the correct one is obtained. Therefore; the required operations for calculating encryption matrix is $\mathcal{O}(KN|Aut(\Gamma)|^K)$. It is clear that if the value of $|Aut(\Gamma)| = 2^{N-K}$ is large enough, then this attack will fail. Furthermore, the average expected number of attempts to obtain all distinct $C_i$ are shown to be $\mathcal{N}_e \sum_{i=0}^{\mathcal{N}_e - 1} 1/(\mathcal{N}_e - i) = \mathcal{O}(\mathcal{N}_e \log \mathcal{N}_e)$. The total number of ciphertexts required for this attack is $\mathcal{O}(K\mathcal{N}_e \log \mathcal{N}_e)$ because this procedure should be repeated for the $K$ unit vectors $\{u_i\}_{i=1}^K$ [4]. The work factor of this attack for the proposed cryptosystem based on $(1024, 768)$ polar codes is approximately $\mathcal{O}(2^{271})$. Also, since the number of error vectors for RN scheme is small, the work factor of ST attack for that scheme is approximately $\mathcal{O}(2^{15})$. Therefore, unlike the RN scheme, the proposed system is secure against the ST attack.

## VI. CONCLUSION

In this paper, we considered possible inclusion of polar codes in a secret key code based cryptosystem. By selecting a set of $K$ secret channel indices $I(\mathcal{S})$ randomly from the $NR_0$ leftmost indices of permutation $\pi_N$, one can construct a large family of equivalent polar codes which helps the proposed cryptosystem resist against the exhaustive search attack. Also, the key length of this scheme is decreased because of storing $K$ secret indices instead of saving generator matrix and using the compression/ decompression algorithms based on circulant block submatrices of $\mathcal{S}$ and $\mathcal{P}$. This cryptosystem is secure against chosen plaintext attacks such as the Rao-Nam and the Struik-Tilburg attacks benefiting from large number of perturbation error vectors and proper choices of code parameters.

ACKNOWLEDGMENT

The authors would like to thank Seyed Hamed Hassani for his helpful discussions.